\documentclass[aps]{revtex4}
\usepackage{amsfonts}
\usepackage{amsmath}
\usepackage{amssymb}
\usepackage{graphicx}

\input{tcilatex}

\begin{document}

\title[Scattering in graphene]{Electrons scattering in the monolayer
graphene with the short-range impurities}
\author{Natalie E. Firsova}
\affiliation{Institute for Problems of Mechanical Engineering, the Russian Academy of
Sciences, St. Petersburg 199178, Russia}
\author{Sergey A. Ktitorov}
\affiliation{A.F. Ioffe Physical-Technical Institute, the Russian Academy of Sciences,
St. Petersburg, Russia}

\begin{abstract}
Scattering problem for electrons in monolayer graphene with short-range
perturbations of the types "local chemical potential" and "local gap" has
been solved. Zero gap and non-zero gap kinds of graphene are considered. The
determined S-matrix can be used for calculation of such observables as
conductance and optical absorption.
\end{abstract}

\maketitle







During the last years much attention was payed to the problem of the
electronic spectrum of graphene (see a review \cite{novosel}).
Two-dimensional structure of it and a presence of the cone points in the
electronic spectrum make actual a comprehensive study of the external fields
effect on the spectrum and other characteristics of the electronic states
described by the Dirac equation in the 2+1 space-time. We consider in this
work the electrons scattering in the 2+1 Dirac equation model of the
monolayer graphene due to the short-range perturbations. We do not take into
account the inter-valley transitions. Particular attention to this case
stems from the effectiveness of short-range scatterers in contrast to the
long-range ones: an effect of the latter is suppressed by the Klein paradox 
\cite{beenakker}. Short-range potential impurities in graphene were
considered in works \cite{basko}, \cite{novikov}, \cite{matulis}. In our
work \cite{we}, a new model of the short-range impurities in graphene was
considered taking into account the obvious fact that the Kohn-Luttinger
matrix elements of the short-range perturbation calculated on the upper and
lower band wave functions are not equal in a general case. This means that
the perturbation must be generically described by a Hermitian matrix. We
considered the diagonal matrix case corresponding to a presence of the
potential and mass perturbation. The bound states dependence on the
perturbation parameters was studied in \cite{we} within the framework of
this model.

In the present work we study the electrons scattering by the short-range
impurities within the framework of the model suggested in \cite{we}.

The Dirac equation describing electronic states in graphene reads 
\begin{equation}
\left( -i\hbar v_{F}\sum_{\mu =1}^{2}\gamma _{\mu }\partial _{\mu }-\gamma
_{0}\left( m+\delta m\right) v_{F}^{2}{}\right) \psi =\left( E-V\right) \psi
,  \label{diracgeneral}
\end{equation}%
where $v_{F}$ is the Fermi velocity of the band electrons, $\gamma _{\mu }$
are the Dirac matrices%
\begin{equation*}
\gamma _{0}=\sigma _{3},\text{ }\gamma _{1}=\sigma _{1},\text{ }\gamma
_{2}=i\sigma _{2},
\end{equation*}%
$\sigma _{i}$ are the Pauli matrices, $2mv_{F}{}^{2}=E_{g}$ is the
electronic bandgap, $\psi \left( \mathbf{r}\right) $ is the two-component
spinor. The electronic gap can appear in the graphene monatomic film lying
on the substrate because of the sublattices mutual shift \cite{gap}. The
spinor structure takes into account the two-sublattice structure of graphene.%
$\ \delta m\left( \mathbf{r}\right) $ and $V(\mathbf{r})$ are the local
perturbations of the mass (gap) and the chemical potential. A local mass
perturbation can be induced by defects in a graphene film or in the
substrate \cite{gap}. We consider here the delta function model of the
perturbation:%
\begin{equation}
\delta m\left( \mathbf{r}\right) =-b\delta (r-r_{0}),\text{ }V(\mathbf{r)}%
=-a\delta (r-r_{0}),  \label{perturb}
\end{equation}%
where $r$ and $r_{0}$ are respectively the polar coordinate radius and the
perturbation radius. Such short-range perturbation was used in the
(3+1)-Dirac problem for narrow-gap and zero-gap semiconductors in \cite%
{tamar}. The perturbation matrix elements 
\begin{equation}
diag(V_{1},V_{2})\delta (r-r_{0})  \label{diag}
\end{equation}%
are related to the $a,$ $b$ parameters as follows 
\begin{equation}
-V_{1}=a+b,\text{ }-V_{2}=a-b  \label{abVrelation}
\end{equation}

The delta function perturbation is the simplest solvable short-range model.
Finite radius $r_{0}$ plays a role of the regulator and is necessary in
order to exclude deep states of the atomic energy scale. The finite
perturbation radius $r_{0}$ leads to the quasi-momentum space form-factor
proportional to the Bessel function that justifies our neglect of
transitions between the Brillouin band points $K$\ and $K^{\prime }$ \cite%
{tamar}.

Let us present the two-component spinor in the form%
\begin{equation}
\psi _{j}(\mathbf{r},t)=\frac{\exp \left( -iEt\right) }{\sqrt{r}}\left( 
\begin{array}{c}
f_{j}\left( r\right) \exp \left[ i\left( j-1/2\right) \varphi \right] \\ 
\\ 
g_{j}\left( r\right) \exp \left[ i\left( j+1/2\right) \varphi \right]%
\end{array}%
\right) ,  \label{spinor}
\end{equation}%
where $j$ is the pseudospin quantum number; $j=\pm 1/2,$ $\pm 3/2,\ldots $.
In opposite to the relativistic theory, this quantum number has nothing to
do with the real spin and indicates a degeneracy in the biconic Dirac point.
The upper $f_{j}\left( r\right) $ and lower $g_{j}\left( r\right) $
components of the spinor satisfy the equations set 
\begin{equation}
\frac{dg_{j}}{dr}+\frac{j}{r}g_{j}-\left( E-m\right) f_{j}=\left( a+b\right)
\delta (r-r_{0})f_{j},  \label{componenteq1}
\end{equation}

\begin{equation}
-\frac{df_{j}}{dr}+\frac{j}{r}f_{j}-\left( E+m\right) g_{j}=\left(
a-b\right) \delta (r-r_{0})g_{j}.  \label{componeq2}
\end{equation}%
These equations have a symmetry: 
\begin{equation}
f_{j}\leftrightarrow g_{j},\text{ }E\rightarrow -E,\text{ }j\rightarrow -j,%
\text{ }a\rightarrow -a.  \label{symm}
\end{equation}%
Let us introduce the function $\varphi _{j}\left( r\right) \equiv
f_{j}/g_{j}.$ It satisfies the equation:%
\begin{equation}
1/\left[ \left( a+b\right) \varphi _{j}^{2}+\left( a-b\right) \right] \left[ 
\frac{d\varphi _{j}}{dr}-\frac{2j}{r}\varphi _{j}-E\left( \varphi
_{j}^{2}+1\right) \right] +\delta (r-r_{0})=0  \label{phi}
\end{equation}%
Integrating in the vicinity of $r=r_{0}$%
\begin{equation}
\lim_{\delta \rightarrow 0}\int_{\varphi _{j}(r_{0}-\delta )}^{\varphi
_{j}(r_{0}+\delta )}\frac{d\varphi _{j}}{\left( a+b\right) \varphi
_{j}^{2}+\left( a-b\right) }=-1,\text{ \ }  \label{match1}
\end{equation}%
we obtain the matching condition \ 

\begin{equation}
\arctan \left( \varphi _{j}^{-}\sqrt{\left( a+b\right) /\left( a-b\right) }%
\right) -\arctan \left( \varphi _{j}^{+}\sqrt{\left( a+b\right) /\left(
a-b\right) }\right) =\sqrt{a^{2}-b^{2}},\text{ \ \ \ }a^{2}>b^{2},
\label{match2}
\end{equation}%
where $\varphi _{j}^{-}\equiv \varphi _{j}\left( r_{0}-\delta \right) ,$ $%
\varphi _{j}^{+}\equiv \varphi _{j}\left( r_{0}+\delta \right) ,$ $\delta
\longrightarrow 0.$ Excluding the spinor component $g_{j}$ from the equation
set Eq. (\ref{componenteq1}), Eq. (\ref{componeq2}) in the domains $0\leq
r<r_{0}$ and $r>r_{0},$ we obtain the second-order equation$:$%
\begin{equation}
\frac{d^{2}f_{j}}{dr^{2}}+\left[ E^{2}-m^{2}-\frac{j\left( j-1\right) }{r^{2}%
}\right] f_{j}=0.  \label{secondorder}
\end{equation}%
This equation is related to the Bessel one. We assume $E$ to be real and
satisfying the inequality $E^{2}\geq m^{2}.$ Then the general solution of
Eq. (\ref{secondorder}) in the region $0\leq r<r_{0}$ reads%
\begin{equation}
f_{j}=C_{1}\sqrt{\kappa r}J_{j-1/2}\left( \kappa r\right) +C_{2}\sqrt{\kappa
r}N_{j-1/2}\left( \kappa r\right) ,  \label{general}
\end{equation}%
where $\kappa =\sqrt{E^{2}-m^{2}}$ is the principal value of the root; $%
J_{\nu }\left( z\right) $ and $N_{\nu }\left( z\right) $ are respectively
the Bessel and Neumann functions. The constant $C_{2\text{ }}$vanishes in
the domain $0\leq r<r_{0}$ since the solution must be regular at the
origin.Expressing the $g_{j}$-component from Eq. (\ref{componeq2}), we can
write%
\begin{equation*}
g_{j}=\sqrt{\frac{E-m}{E+m}}\sqrt{\kappa r}C_{1}J_{j+1/2}\left( \kappa
r\right) .
\end{equation*}
Thus 
\begin{equation}
\varphi _{j}^{-}\left( \kappa r_{0}\right) =\sqrt{\frac{E+m}{E-m}}\frac{%
J_{j-1/2}\left( \kappa r_{0}\right) }{J_{j+1/2}\left( \kappa r_{0}\right) }.
\label{phi-}
\end{equation}%
Then we can obtain from Eq. (\ref{match2}):

\begin{equation}
\arctan \left( \sqrt{\frac{a+b}{a-b}}\varphi _{j}^{+}\left( \kappa
r_{0}\right) \right) =\arctan \left( \sqrt{\frac{a+b}{a-b}}\sqrt{\frac{E+m}{%
E-m}}\frac{J_{j-1/2}\left( \kappa r_{0}\right) }{J_{j+1/2}\left( \kappa
r_{0}\right) }\right) -\sqrt{a^{2}-b^{2}},  \label{match3}
\end{equation}%
and, therefore,%
\begin{equation}
\varphi _{j}^{+}\left( \kappa r_{0}\right) =\frac{\sqrt{\frac{E+m}{E-m}}%
J_{j-1/2}\left( \kappa r_{0}\right) -\left( a-b\right) T\left( a,b\right)
J_{j+1/2}\left( \kappa r_{0}\right) }{J_{j+1/2}\left( \kappa r_{0}\right)
+\left( a+b\right) \sqrt{\frac{E+m}{E-m}}T\left( a,b\right) J_{j-1/2}\left(
\kappa r_{0}\right) },  \label{phi+2}
\end{equation}%
where $T\left( a,b\right) $ is given by the formula:%
\begin{equation}
T\left( a,b\right) =\left\{ 
\begin{tabular}{l}
$\frac{\tan \left( \sqrt{a^{2}-b^{2}}\right) }{\sqrt{a^{2}-b^{2}}}$ \ \ if $%
a^{2}>b^{2},$ \\ 
\\ 
$\frac{\tanh \left( \sqrt{b^{2}-a^{2}}\right) }{\sqrt{b^{2}-a^{2}}}$ \ \ \
if $b^{2}>a^{2},$%
\end{tabular}%
\right. .  \label{T}
\end{equation}%
On the other hand, an expression for $\varphi _{j}^{+}\left( \kappa
r_{0}\right) $ can be written similarly to \ref{phi-}:

\begin{equation}
\varphi _{j}^{+}\left( \kappa r_{0}\right) =\frac{f_{j}^{+}}{g_{j}^{+}}=%
\sqrt{\frac{E+m}{E-m}}\frac{H_{j-1/2}^{\left( 2\right) }\left( \kappa
r_{0}\right) +S_{j}H_{j-1/2}^{\left( 1\right) }\left( \kappa r_{0}\right) }{%
H_{j+1/2}^{\left( 2\right) }\left( \kappa r_{0}\right)
+S_{j}H_{j+1/2}^{\left( 1\right) }\left( \kappa r_{0}\right) },
\label{phi+3}
\end{equation}%
where $S_{j}\left( \kappa \right) $ is a phase factor of the out-going wave,
i. e. S-matrix element in the angular momentum representation. Substituting
Eq. (\ref{phi+3}) into Eq. (\ref{phi+2}), we obtain an explicit expression
for $S_{j}\left( E\right) $: 
\begin{equation}
S_{j}\left( E\right) =-\frac{\mathcal{F}_{j}^{\left( 2\right) }}{\mathcal{F}%
_{j}^{\left( 1\right) }},  \label{S2}
\end{equation}%
where

\begin{eqnarray}
\mathcal{F}_{j}^{\left( \alpha \right) } &=&\left( J_{j-1/2}\left( \kappa
r_{0}\right) H_{j+1/2}^{\left( \alpha \right) }\left( \kappa r_{0}\right)
-J_{j+1/2}\left( \kappa r_{0}\right) H_{j-1/2}^{\left( \alpha \right)
}\left( \kappa r_{0}\right) \right) -  \notag \\
&&T\left( a,b\right) \left[ \sqrt{\frac{E-m}{E+m}}\left( a-b\right)
J_{j+1/2}\left( \kappa r_{0}\right) H_{j+1/2}^{\left( \alpha \right) }\left(
\kappa r_{0}\right) +\sqrt{\frac{E+m}{E-m}}\left( a+b\right) J_{j-1/2}\left(
\kappa r_{0}\right) H_{j-1/2}^{\left( \alpha \right) }\left( \kappa
r_{0}\right) \right] .  \label{F}
\end{eqnarray}%
Here $\alpha $ takes values $0,1.$ Since $H_{n}^{\left( 2\right) }\left(
z\right) =H_{n}^{\left( 1\right) \ast }\left( z\right) $ for real $z,$ the
scattering matrix is unitary everywhere on the continuum spectrum. Eq. (\ref%
{S2}) solves the electron scattering problem for the given potential. The
denominator of $S_{j}(E)$ is just the left-hand side of the characteristic
equation derived in \cite{we}. Imaginary roots of it correspond to the real
energy eigenstates (bound states) lying in thegap, which were studied in
that paper. The characteristic equation reads%
\begin{equation}
\mathcal{F}_{j}^{\left( 1\right) }\left( \kappa r_{0}\right) =0,
\label{char}
\end{equation}%
or%
\begin{eqnarray}
&&\left( J_{j-1/2}\left( \kappa r_{0}\right) H_{j+1/2}^{\left( \alpha
\right) }\left( \kappa r_{0}\right) -J_{j+1/2}\left( \kappa r_{0}\right)
H_{j-1/2}^{\left( \alpha \right) }\left( \kappa r_{0}\right) \right)   \notag
\\
&=&T\left( a,b\right) \left[ \sqrt{\frac{E-m}{E+m}}\left( a-b\right)
J_{j+1/2}\left( \kappa r_{0}\right) H_{j+1/2}^{\left( \alpha \right) }\left(
\kappa r_{0}\right) +\sqrt{\frac{E+m}{E-m}}\left( a+b\right) J_{j-1/2}\left(
\kappa r_{0}\right) H_{j-1/2}^{\left( \alpha \right) }\left( \kappa
r_{0}\right) \right]   \label{char2}
\end{eqnarray}%
Using the relations $H_{n}^{\left( 1\right) }\left( z\right) =J_{n}+iN_{n},$ 
$H_{n}^{\left( 2\right) }=J_{n}-iN_{n},$ we can write S-matrix in the form:%
\begin{equation}
S_{j}\left( E\right) =-\frac{A_{j}\left( E\right) +iB_{j}\left( E\right) }{%
A_{j}\left( E\right) -iB_{j}\left( E\right) }=\frac{B_{j}\left( E\right)
+iA_{j}\left( E\right) }{B_{j}\left( E\right) -iA_{j}\left( E\right) },
\label{swrational}
\end{equation}%
and, therefore, it can be presented in the standard form \cite{KMLL}%
\begin{equation}
S_{j}\left( E\right) =\exp \left[ i2\delta _{j}\left( E\right) \right] ,
\label{phase}
\end{equation}%
where the scattering phase is given by the expression 
\begin{equation}
\delta _{j}\left( E\right) =\arctan \frac{A_{j}\left( E\right) }{B_{j}\left(
E\right) }.  \label{delta}
\end{equation}%
Formulae (\ref{swrational}), (\ref{phase}) show that the scattering matrix $%
S_{j}\left( E\right) $ is unitary on the continuum spectrum. The functions $%
A_{j}\left( E\right) $ and $B_{j}\left( E\right) $ are determined as follows%
\begin{equation}
A_{j}\left( E\right) =-T\left( a,b\right) \left[ \left( a+b\right) \sqrt{%
\frac{E+m}{E-m}}J_{j-1/2}^{2}\left( \kappa r_{0}\right) +\left( a-b\right) 
\sqrt{\frac{E-m}{E+m}}J_{j+1/2}^{2}\left( \kappa r_{0}\right) \right] ,
\label{A}
\end{equation}

\begin{eqnarray}
B_{j}\left( E\right) &=&T\left( a,b\right) \left[ \left( a+b\right) \left( 
\sqrt{\frac{E+m}{E-m}}\right) J_{j-1/2}\left( \kappa r_{0}\right)
N_{j-1/2}\left( \kappa r_{0}\right) +\left( a-b\right) \sqrt{\frac{E-m}{E+m}}%
J_{j+1/2}\left( \kappa r_{0}\right) N_{j+1/2}\left( \kappa r_{0}\right) %
\right] +  \notag \\
&&\left[ J_{j+1/2}\left( \kappa r_{0}\right) N_{j-1/2}\left( \kappa
r_{0}\right) -J_{j-1/2}\left( \kappa r_{0}\right) N_{j+1/2}\left( \kappa
r_{0}\right) \right]  \label{B}
\end{eqnarray}%
It is seen from (\ref{delta}), (\ref{B}) that all $\delta _{j}\left(
E\right) $ vanish , when $a$ and $b$ tend to zero, i. e. in the absence of a
perturbation. Using the Bessel functions expansion \cite{stegun} 
\begin{equation}
J_{n}\left( x\right) \sim \left( 1/n!\right) \left( x/2\right) ^{n},
\label{J}
\end{equation}%
\begin{equation}
N_{n}\left( x\right) \sim \left\{ 
\begin{array}{c}
-\left( \Gamma \left( n\right) /\pi \right) \left( 2/x\right) ^{n}\text{ \ \
\ for }n>0, \\ 
\\ 
\left( 2/\pi \right) \log \left( \gamma _{E}x/2\right) \text{ \ \ \ for }n=0%
\end{array}%
\right.  \label{N}
\end{equation}%
we conclude that for the low-energy scattering $\kappa r_{0}<<1,$ $\delta
_{j}\left( E\right) $ is small as $\left( \kappa r_{0}\right) ^{\left\vert
j\right\vert +1/2}$ except of $j=\pm 1/2.$ Here $\log \gamma _{E}$ is the
Eyler-Mascheroni constant. In the case of small radius $r_{0}$ and low
energy $E$ we can neglect all higher angular momentum partial waves taking
into account only phases $\delta _{j}$ for $j=\pm 1/2:$

\begin{equation*}
\tan \delta _{1/2}\left( E\right) =
\end{equation*}

\begin{equation*}
=-T\left( a,b\right) \frac{\left( a+b\right) \sqrt{\frac{E+m}{E-m}}+\left(
a-b\right) \sqrt{\frac{E-m}{E+m}}\left( \kappa r_{0}/2\right) ^{2}}{\left[
\left( \kappa r_{0}/2\right) \frac{2}{\pi }\log \left( \gamma _{E}\kappa
r_{0}/2\right) -\frac{1}{\pi }\left( 2/\kappa r_{0}\right) \right] +T\left(
a,b\right) \left[ \left( a+b\right) \sqrt{\frac{E+m}{E-m}}\frac{2}{\pi }\log
\left( \gamma _{E}\kappa r_{0}/2\right) +\left( a-b\right) \sqrt{\frac{E-m}{%
E+m}}\frac{\Gamma \left( 1\right) }{\pi }\right] }\approx
\end{equation*}

\begin{equation}
T\left( a,b\right) \sqrt{\frac{E+m}{E-m}}\left( a+b\right) \pi \left( \frac{%
\kappa r_{0}}{2}\right) ,\text{ }\kappa r_{0}\longrightarrow 0
\label{phase+1/2}
\end{equation}

\begin{equation*}
\tan \delta _{-1/2}\left( E\right) =
\end{equation*}

\begin{equation*}
=-T\left( a,b\right) \frac{\left( a-b\right) \sqrt{\frac{E-m}{E+m}}+\left(
a+b\right) \sqrt{\frac{E+m}{E-m}}\left( \varkappa r_{0}/2\right) ^{2}}{\left[
\frac{\Gamma \left( 1\right) }{\pi }\left( 2/\kappa r_{0}\right) -\left(
\kappa r_{0}/2\right) \frac{2}{\pi }\log \left( \gamma _{E}\kappa
r_{0}/2\right) \right] +T\left( a,b\right) \left[ \sqrt{\frac{E-m}{E+m}}%
\left( a-b\right) \frac{2}{\pi }\log \left( \gamma _{E}\kappa r_{0}/2\right)
+\sqrt{\frac{E+m}{E-m}}\left( a+b\right) \frac{\Gamma \left( 1\right) }{\pi }%
\right] }\approx
\end{equation*}

\begin{equation}
-T\left( a,b\right) \sqrt{\frac{E-m}{E+m}}\left( a-b\right) \pi \left( \frac{%
\kappa r_{0}}{2}\right) ,\text{ \ \ }\kappa r_{0}\longrightarrow 0
\label{phase-1/2}
\end{equation}

\bigskip We see that the phase is proportional to $\kappa r_{0}$ in the
long-wave limit as it is necessary \cite{KMLL}, \cite{novikov}. The
scattering amplitude $f\left( \theta \right) $ and transport cross-section $%
\Sigma _{tr}$ can be expressed in terms of $S_{j}\left( E\right) $ as follow 
\cite{novikov}:%
\begin{equation}
f\left( \theta \right) =\frac{1}{i\sqrt{2\pi \kappa }}\sum_{j=\pm 1/2,\pm
3/2,...}\left[ S_{j}\left( E\right) -1\right] \exp \left[ i\left(
j-1/2\right) \theta \right] ,  \label{amplitude}
\end{equation}%
\begin{equation}
\Sigma _{tr}=2/\kappa \sum_{j=\pm 1/2,\pm 3/2,..}\sin ^{2}\left( \delta
_{j+1}-\delta _{j}\right)   \label{crosssection}
\end{equation}%
Near the resonance states the Breit-Wigner form of the phase is valid \cite%
{KMLL}: 
\begin{equation*}
\delta _{j}\approx \delta _{j}^{\left( 0\right) }+\arctan \frac{\Gamma _{j}}{%
2\left( E_{j}^{(0)}-E\right) },
\end{equation*}%
where $E_{j}^{(0)}$ and $\Gamma _{j}$ are respectively the position and
width of the resonance level, $\delta _{j}^{\left( 0\right) }$ is the
slowly-varying potential scattering phase.

The presented above formulae can be used in order to calculate the Boltzmann
conductivity \cite{sarma}

\begin{equation}
\sigma =\left( \frac{e^{2}}{2\pi \hslash }\right) \frac{2E_{F}}{\hslash }%
\tau _{tr},  \label{conductivity.}
\end{equation}%
where the transport relaxation time equals

\begin{equation}
1/\tau _{tr}=N_{i}v_{F}\Sigma _{tr}.  \label{relaxtime}
\end{equation}%
Here $N_{i}$ is the areal impurity density, $E_{F}=v_{F}\kappa _{F}$. The
above equations transform a dependence of the scattering data on the Fermi
energy and impurity perturbation parameters $a$ and $b$ into the
correspondent dependence of the Boltzmann conductivity. Thus characteristic
features of the scattering data determine a behaviour of the electric
conductivity. Proper numeric calculations will be presented elsewhere.

\end{document}